\documentstyle[twoside,fleqn,espcrc2,psfig]{article}

   \newcommand{\be}{\begin{equation}}
   \newcommand{\ee}{\end{equation}}
   
\newcommand{\AmS}{{\protect\the\textfont2
  A\kern-.1667em\lower.5ex\hbox{M}\kern-.125emS}}

\title{Strongly Coupled QCD at Finite Baryon Density}

\author{R. Aloisio\address{Dipartimento di Fisica, Universit\'a di L'Aquila, 
        via Vetoio, 67100 L'Aquila, Italy}$^{,d}$\thanks{Poster presented 
	by R. Aloisio},
	V. Azcoiti\address{Departamento de F\'\i sica Te\'orica,
	Facultad de Ciencias,Universidad de Zaragoza,50009 Zaragoza,Spain}
	G. Di Carlo\address{Istituto Nazionale di Fisica Nucleare,
	Laboratori Nazionali di Frascati, P.O.B. 13,
	00044 Frascati, Italy},
	A. Galante$^b$,
	A.F. Grillo\address{Istituto Nazionale di Fisica Nucleare,
	Laboratori Nazionali del Gran Sasso, 
	67010 Assergi, Italy}}
 
\begin{document}
\pagestyle{empty}
\begin{abstract}
The analytical results obtained in the infinite mass and strong coupling limits
of QCD are difficult to reconcile with the predictions of the Monomer 
Dimer Polymer algorithm. We have reconsidered in detail the results obtained 
with this simulation scheme and evidences of severe convergence problems
are presented for the $SU(3)$ and $SU(2)$ gauge group cases.
\end{abstract}

\maketitle
\section*{Introduction}

Finite Density QCD is affected by the well known sign problem 
that has prevented, in most cases, any success in simulate this theory and, 
until now, no solution is at sight. 
The only exceptions in this scenario are strong coupling simulations 
performed with the Monomer Dimer Polymer (MDP) algorithm \cite{MDP1}.
This algorithm is able to provide results not affected by early onset 
and in partial agreement with the Mean Field (MF) prediction.
More recently another theoretical advance has been achieved solving 
exactly the $\beta=0$ QCD in the limit of infinite mass and chemical 
potential \cite{massa}. 
In the next section we discuss the compatibility of MDP results with the
infinite mass $SU(3)$ solution; indeed we have found difficult to 
conciliate the numerical and analytical predictions. This leds us to reconsider
the MDP algorithm more carefully and we have found evidences of convergence 
problems.

We have also considered the $SU(2)$ case where the sign problem 
is not present. Results obtained using the Gran Canonical Partition
Function (GCPF) formalism turn out to be in very good agreement with 
the Hybrid Montecarlo (HMC) calculations \cite{han} while, once again,
MDP results \cite{MDP2} are inaccurate in the critical region.

\section*{$SU(3)$ case}

The main feature of infinite mass QCD \cite{massa} is that
at zero temperature $(L_t\to\infty)$ the system undergoes a first order
saturation transition and the phase of the Dirac determinant is not relevant.
However at non-zero temperature ($L_t$ finite) the system has only a
smooth crossover and the phase is relevant.
The MDP results are somehow puzzling if considered in the light 
of the infinite mass solution. In \cite{MDP1} Karsch and M\"utter 
saw a strong first order transition for $L_t=4$ and $L_s=4,8$
with $m\in [0.1,0.7]$, while we know that for $m\to\infty$ the number
density is a smooth function of $\mu$ for any finite $L_t$.
To reconcile the $m\to\infty$ solution and MDP results we have to suppose that 
the transition disappears at some large bare mass $\bar{m}$, or that the 
infinite mass limit of QCD is singular.
These statements seem both unnatural. The former would imply the
existence of a (large) physical scale where the system behaviour changes,
washing out the transition. The latter is unplausible too, because the Dirac 
determinant approaches its infinite mass limit continuously.
In order to have data more easily comparable with the analytic predictions 
we have tried to use the MDP algorithm
\footnote{The authors thank F. Karsch for providing the MDP code}
directly in the large mass regime. 
The authors of the MDP code noticed
\cite{MDP1} that for small masses (i.e. $m<0.1$)  
the algorithm becomes non-ergodic. When we tried MDP simulations for $m>1.0$ we 
saw a similar behaviour: for no value of $\mu$ the system 
moved into the saturated phase.
Moreover a degradation in performance (for all mass values) has been observed 
increasing the lattice size but, what is more surprising, this 
degradation seems to be related only to the value of $L_t$. We have not been 
able to perform simulations for $L_t>4$.

The behaviour of MDP code prevents any direct comparison between the
large mass results and MDP ones. The limited applicability of MDP
algorithm raises doubts on its general validity.
We have repeated the MDP simulations at $m=0.1$ in a 
$4^3\times 4$ lattice (values used in the original paper \cite{MDP1}),
using as initial configuration either $n(\mu)=0$ or $n(\mu)=1$ and
$O(10^6)$ Montecarlo steps for each value of $\mu$. 
\begin{figure}[!t]
\psrotatefirst
\psfig{figure=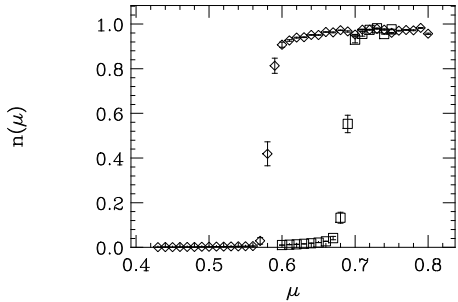,angle=90,width=200pt}
\caption{Number density as function of $\mu$ with:
$\mu=0$ start (diamonds), saturated start (squares).}
\label{fig:1}
\end{figure}
From fig. 1 it is evident a signal of hysteresis in the 
data; the $\mu=0$ start undergoes a strong saturation transition at 
$\mu=0.69$ (published result \cite{MDP1}),
while the run with saturated start jumps in the zero 
density phase at $\mu=0.58$. This result should lead us to cautioun on the
determination of the critical point and may well reconcile the MDP results
with MF predictions (the MF critical point $\mu_c^{MF}=0.61$ lies inside the
hysteresis). 
To be confident with MDP results we should observe several flip-flops in the 
Montecarlo history to conclude that a clear two state signal is present.
We have considered runs of up to $O(10^9)$ configurations and
we did not succeed to see any flip-flop for any $\mu$ inside the 
hysteresis region.
Starting from a zero density configuration nothing happens until we get 
close to $\mu=0.69$. At this point the system has some probability to go 
in the saturated phase. Once the system is in the saturated phase it never 
goes back. The same behaviour has been observed (near $\mu=0.58$) for 
the saturated start. Varying the quark mass only changes the hysteresis 
position unless we reach too small $(m<0.1)$ or too large $(m>1.0)$
values. 
From these numerical evidences we may conclude that the hysteresis behaviour 
of the system is independent on the statistics for any value of $\mu$.
The MDP code seems to have convergence problems in the 
most interesting region of $\mu$, independently on the temporal lattice extent 
and the quark mass value. 

\section*{$SU(2)$ case}

Let us now address the MDP convergence in the case of $SU(2)$
gauge group. The motivation is twofold: firstly we wonder if the problems 
present in $SU(3)$ are universal independently on the gauge group. Moreover
$SU(2)$ offers us the possibility of using conventional simulation algorithms.
In this case, in fact, quarks and antiquarks belong to the same (real) 
representation: the Dirac determinant is real and positive also for 
non-zero $\mu$.
The main published MDP results in $SU(2)$ concern the number density and the
chiral condensate in a $4^3\times 4$ and $8^3\times 4$ lattice at $m=0.2$
\cite{MDP2}. Other results for $SU(2)$ at $\beta=0$ and 
$\mu\ne 0$ are obtained using the HMC algorithm in a $4^3 \times 4$
at $m=0.2$ \cite{han}. In order to have results with $\mu$ varying 
continuously, we have used the GCPF scheme, performing simulations with 
$L_t=4$ and $L_s=4,6,8$, at the quark mass values $m=0.1,0.2,0.4$. 
\begin{figure}[!t]
\psrotatefirst
\psfig{figure=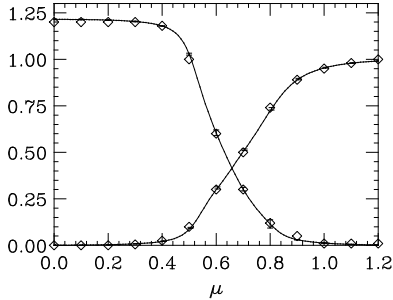,angle=90,width=200pt}
\caption{ Number density and chiral condensate as a function of $\mu$ 
in a $4^3 \times 4$ at $m=0.2$ from GCPF (line) and HMC (symbols).}
\label{fig:5}
\end{figure}
In fig. 2 we report the number density and the chiral condensate
obtained in our simulations compared with HMC results appeared in 
fig. 4 of \cite{han}; error bars are reported at some values of $\mu$. 
It is evident that our simulation reproduces the HMC 
results quite accurately.
This agreement between GCPF and HMC suggests that sampling problems,
found in $SU(3)$ with GCPF \cite{massa}, are not present in $SU(2)$. 
Simplified models predict for $SU(2)$, at least at small temperature,
a phase transition at half of the mass of the lightest baryon of the theory 
(degenerated with the pion at $\mu=0$). We have then computed the pion mass
in a $6^3\times 12$ at $m=0.1,0.2,0.4$.
To extract the critical $\mu$ we have used the following criterium. 
The number density appears, increasing the volume, almost zero up to 
$\mu_c$, with a linear rise beyond it and flat at large $\mu$ (saturation). 
To identify the critical point we have computed 
$\partial n(\mu)/\partial\mu$ for two different volumes and defined $\mu_c$ 
as the position of the first crossing of the curves. In the 
limit $V\to\infty$ this defines correctly the $\mu$ at which the linear 
behaviour begins. In table we report our $\mu_c$ and one half the pion 
mass for different values of $m$:
\begin{center}
\begin{tabular}{|c|c|c|}
\ $m$ & $\mu_c$  & $\frac{m_{\pi}}{2}$ \\
\hline
\ 0.1 & 0.340(4) & 0.3408(7) \\
\ 0.2 & 0.485(5) & 0.4840(6) \\
\ 0.4 & 0.693(5) & 0.6889(5) \\
\end{tabular}
\end{center}
We can conclude from these data that our predicted $\mu_c$ moves 
with the quark mass in the expected way; this behaviour increases our
confidence on our results.
\begin{figure}[!t]
\psrotatefirst
\psfig{figure=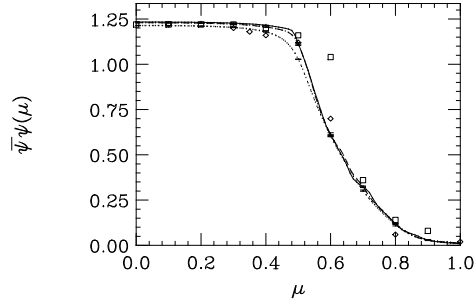,angle=90,width=200pt}
\caption{Chiral condensate at $m=0.2$ in $4^3 \times 4$ 
(dots), $6^3 \times 4$ (dashes), $8^3 \times 4$ (line) 
and MDP results in a $4^3 \times 4$ (diamonds) and $8^3 \times 4$ 
(squares).}
\label{fig:7}
\end{figure}
In fig. 3 we report the chiral condensate 
at m=0.2 from our data and, superimposed, the MDP results as reported
in figure 2 of \cite{MDP2}.
From fig. 3 it is evident a marked difference between MDP results and those
by our simulations only for the largest lattice $(8^3 \times 4)$ and again 
limited around $\mu_c$ as in the $SU(3)$ case. In particular the 
$8^3 \times 4$ results differ from ours at $\mu=0.6$, the critical point 
derived in \cite{MDP2}. The authors of \cite{MDP2} have tested the 
independence of their results on the initial configuration only for the 
$4^3 \times 4$ lattice. In this case MDP results agree at a good level with
ours (see $\bar{\psi}\psi$ of fig. 3). In our opinion the observed 
discrepancy has to be ascribed to convergence problems of the MDP algorithm,
although they arise at volumes larger then in the $SU(3)$ case. Once again
there are serious doubts on the accuracy that the MDP algorithm can achieve
near the critical region.

\end{document}